\documentclass[aps,pra,twocolumn,superscriptaddress,showkeys,a4paper]{revtex4-2}
\usepackage{amsfonts}
\usepackage{amssymb}
\usepackage{mathrsfs}
\usepackage[intlimits]{amsmath}
\usepackage{bm, bbm}
\usepackage[dvips]{graphicx}
\usepackage[T1]{fontenc}
\usepackage{amsmath}

\usepackage{latexsym}
\usepackage[makeroom]{cancel}

\usepackage{color}

\def\dt{{\rm d}\,}

\newcommand{\ket}[1]{| #1 \rangle}
\newcommand{\bra}[1]{\langle #1 |}

\def\duzomniejsze{<\kern-.7mm<}
\def\duzowieksze{>\kern-.7mm>}

\def\textbf#1{{\bf #1}}
\def\be{\begin{equation}}
\def\ee{\end{equation}}
\def\ben{\begin{eqnarray}}
\def\een{\end{eqnarray}}
 \def\beqa{\begin{eqnarray}}
\def\eeqa{\end{eqnarray}}
\def\eea{\end{array}}
\def\bea{\begin{array}}
\newcommand{\bei}{\begin{itemize}}
\newcommand{\eei}{\end{itemize}}
\newcommand{\bee}{\begin{enumerate}}
\newcommand{\eee}{\end{enumerate}}

\def\1{\openone}

\def\tr{{\rm Tr}}

\def\>{\rangle}
\def\<{\langle}

\def\dt#1{{{\kern -.0mm\rm d}}#1\,}
\def\squareforqed{\hbox{\rlap{$\sqcap$}$\sqcup$}}
\def\qed{\ifmmode\squareforqed\else{\unskip\nobreak\hfil
\penalty50\hskip1em\null\nobreak\hfil\squareforqed
\parfillskip=0pt\finalhyphendemerits=0\endgraf}\fi}

\newtheorem{lemma}{Lemma}
\newtheorem{theorem}[lemma]{Theorem}
\newtheorem{main result}[lemma]{Main result}
\newtheorem{proposition}[lemma]{Proposition}
\newtheorem{definition}{Definition}

\newtheorem{fact}[lemma]{Fact}

\def\bep{\begin{proposition}}
\def\eep{\end{proposition}}
\def\bel{\begin{lemma}}
\def\eel{\end{lemma}}

\def\bet{\begin{theorem}}
\def\eet{\end{theorem}}
\def\bed{\begin{definition}}
\def\eed{\end{definition}}
\def\bef{\begin{fact}}
\def\eef{\end{fact}}

\begin{document}

\title{Complementarity between decoherence and information retrieval from the environment}

\author{Tae-Hun Lee}
\affiliation{Center for Theoretical Physics, Polish Academy of Sciences,\\ Aleja Lotnik\'ow 32/46, 02-668 Warsaw, Poland}
\author{Jaros\l{}aw K. Korbicz}\email{jkorbicz@cft.edu.pl}
\affiliation{Center for Theoretical Physics, Polish Academy of Sciences,\\ Aleja Lotnik\'ow 32/46, 02-668 Warsaw, Poland}
\date{\today}

\begin{abstract}
We address the  problem of fundamental limitations of information extraction from the environment in open quantum systems. We derive a model-independent, hybrid quantum-classical solution of open dynamics in the recoil-less limit, which includes environmental degrees of freedom.  Specifying to the celebrated Caldeira-Leggett model of hot thermal environments, ubiquitous in everyday situations, we reveal the existence of a new  lengthscale, called distinguishability length, different from the well-known thermal de Broglie wavelength that governs the decoherence.  Interestingly, a new integral kernel, called Quantum Fisher Information kernel, appears in the analysis. It complements the well-known dissipation and noise kernels and satisfies  disturbance-information gain type of relations, similar to the famous fluctuation-dissipation relation.  Our results complement the existing treatments of the Caldeira-Legget model from a non-standard and highly non-trivial perspective of  information dynamics in the environment. This leads to a full picture of how the open evolution looks like from both the system and the environment points of view, as well as sets limits on the precision of indirect observations.

\end{abstract}


\maketitle

\section{Introduction}
One of the perpetual questions is if what we perceive is really "out there"? While ontology of quantum mechanics is still a matter of a debate (see e.g.  \cite{Lewis2016, Maudlin2019, Barrett2019}),  it is nowadays commonly accepted following the seminal works of Zeh \cite{Zeh1970} and Zurek \cite{Zurek1981, Zurek1982} that interactions with the environment and resulting decoherence processes  lead to an effective emergence of classical properties, like position \cite{Breuer2002, Joos2003, Schlosshauer2007} . 
It is then usually argued, using idealized pure-state environments,  that the decoherence efficiency corresponds directly to the amount of information recorded by the environment (see e.g. \cite{Schlosshauer2007}). The more environment learns about the system, the stronger it decoheres it. On the other hand, we perceive the outer world  by observing the environment and
the information content of the latter determines what we see.

Here we show that there is a gap between the two: What the environment learns about the system, as determined by the decohering power, and what can be extracted from it via measurements. Some part of information stays bounded.
We show it for physically most relevant  thermal environments in the Caldeira-Leggett regime \cite{Caldeira1983}, which is the universal choice for high temperature environments, ubiquitous in real-life situations. One of the most emblematic results of the whole decoherence theory states that spatial coherences decay on the lengthscale given by the thermal de Broglie wavelength $\lambda_{dB}$ and on the timescale $t_{dec}\sim\gamma^{-1}(\lambda_{dB}/d)^2$, where $d$ is a separation and $\gamma^{-1}$ is related to the relaxation time \cite{Papa1968, Caldeira1983, Moore1986}. 
We complement this celebrated result by analyzing information extractable from the environment as quantified by the state distinguishability \cite{Fuchs1999}. We show that it is governed by a new lengthscale, which we call distinguishability length, 
larger than the decoherence length. Thus the resolution with which system's position can be read-off from the environment is worse than the decoherence resolution. 
A part of information gained by the environment during the decoherence is bounded in it in the thermal noise, similarly to e.g. bounding a part of thermodynamic energy as thermal energy, unavailable for work, or to bounding quantum entanglement \cite{Horodecki1998}.
We obtain the corresponding distinguishability timescale and introduce a new integral kernel, Quantum Fisher Information (QFI) kernel, similar to the well-known noise and dissipation  kernels \cite{Breuer2002, Joos2003, Schlosshauer2007} and governing the distinguishability process.
The discovery of this phenomenon was possible due to the paradigm change in studies of open quantum systems initiated by quantum Darwinism \cite{Zurek2009, Blume2008, Brandao2015, Unden2019, Lorenzo2020, Touil2022} and Spectrum Broadcast Structures (SBS) \cite{Korbicz2014, Horodecki2015, Tuziemski2015, Le2019, Korbicz2021} programs.
They recognize the environment as a carrier of useful information about the system, rather than just the source of noise and dissipation, and study its information content in the context of the quantum-to-classical transition.
The existence of the gap can be in principle deduced from the existing literature on quantum Darwinism, e.g. \cite{Touil2022},
and the corresponding timescale separation was shown in e.g. \cite{Korbicz2017}. However, those studies were perfomed in finite-dimensional settings. Here we study a continous variable system, which called for new methods.
Some hints on the effect were also obtained in earlier studies of SBS, especially in the Quantum Brownian Motion (QBM) model \cite{Tuziemski2015}, where state distinguishability and its temperature dependence was analyzed, but the limited, numerical character of the studies did not reveal
the existence of the distinguishability length and the gap to the decoherence lengthscale. Our results, apart from showing intrinsic limitations of indirect observations, also characterize decoherence,  which has become one of the key paradigms of modern quantum science \cite{Chiorescu2003, Deleglise2008, Haffner2008, Hornberger2012, Moser2014, Tighineanu2018, Fein2019}, from a little studied perspective of the ``receiver's end''.  Last but not least, we uncover an interesting new feature  of the venerable Caldeira-Leggett  model

\begin{figure}[t]
	\centering
	\includegraphics[width=0.49\textwidth]{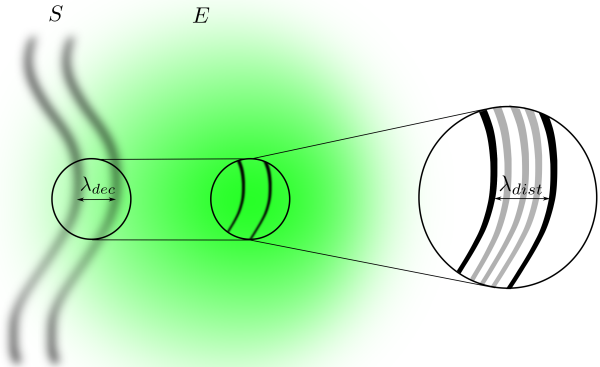}
	\caption{The environment decoheres the central system at a lengthscale $\lambda_{dec}$ (equal to the thermal de Broglie wavelength $\lambda_{dB}$ in the studied example) as a result of dynamical  build-up of correlations  and information leakage into the environment. However not all of that information is accessible, the retrieval is limited by its own resolution, the distinguishability length $\lambda_{dist}$.
Since $\lambda_{dist} \gg \lambda_{dec}$, part of the information stays bounded in the environment. 
Decoherence and distinguishability are complementary to each other as reflected by information gain vs. disturbance type of a relation: $\lambda_{dist}\lambda_{dec} =$ const.
The accompanying  timescales  satisfy $t_{dist} /t_{dec} \sim (\lambda_{dist}/\lambda_{dec})^2$, so that reaching a given information retrieval resolution takes much longer time than reaching the same decoherence resolution.
}
	\label{fig1}
\end{figure}

Although there exist powerful methods of analysis of quantum open systems, such as the Bloch-Redfield \cite{Redfield1957, Bloch1957} or Davies-Gorini-Kossakowski-Lindblad-Sudarshan \cite{Davies1974, Gorini1976, Lindblad1976} 
equations, they describe the evolution of the central system alone, neglecting the environment completely.
This will not tell anything about information acquired by the environment and we instead derive an approximate solution method focusing on the evolution of the environment. To this end, we divide the environment into 
two parts, one denoted $E_{uno}$ is assumed to be unobserved and hence traced over, while the remaining part, denoted $E_{obs}$ is assumed to be under observation by an external observer. Our main object of study will be, so called, partially traced state, obtained by tracing out only the unobserved part of the environment \cite{Korbicz2014, Horodecki2015, Tuziemski2015, Korbicz2021}:  
\be
\varrho_{S:E_{obs}}=\tr_{E_{uno}}\varrho_{S:E},\label{rseobs}
\ee 
Here $S:E_{obs}$ denotes that the resulting state is a joint state of the system $S$ and the observed part of the environment $E_{obs}$.
As the first approximation, it is enough to consider the recoiless limit, where the central system $S$ influences the environment $E$
but is massive enough not to feel the recoil. It is a version of the Born-Oppenheimer approximation and the opposite, and less studied, limit to the commonly used Born-Markov approximation \cite{Breuer2002, Joos2003, Schlosshauer2007}, where the influence $S\to E$ is completely cut out and is thus useless for our purposes. 

\section{Hybrid quantum-classical dynamics in the recoil-less limit}
We follow the  treatment of Feynman and Vernon \cite{Feynman1963} using path integrals. The full System-Environment propagator reads:
\begin{eqnarray}
&&K_t(X,X_0;x,x_0)=\int_{\begin{subarray}{l} x(0)=x_0 \\ x(t)=x \end{subarray}}\mathscr Dx(t) \int_{\begin{subarray}{l} X(0)=X_0 \\ X(t)=X \end{subarray}}\mathscr DX(t)\label{K}\\
&&\times \text{exp}\frac{i}{\hbar}\left\{S_{sys}[X(t)]+S_{env}[x(t)]+S_{int}[X(t),x(t)]\right\}\nonumber,
\end{eqnarray}
where $S_{sys}, S_{env}, S_{int}$ are the actions of the system, environment and interaction respectively; 
$X(t)$ is the system trajectory with the initial condition $X(0)=X_0$, similarly $x(t)$ is the environment trajectory with $x(0)=x_0$.
For a massive enough central system we may neglect the recoil of the environment:
\be\label{nbr}
\left|\frac{\delta S_{sys}}{\delta X(t)}[X(t)]\right|\gg \left|\frac{\delta S_{int}}{\delta X(t)}[X(t),x(t)]\right|.
\ee
and expand the parts containing $X(t)$ around a classical trajectory $X_{cl}(t;X_0)$ 
(in what follows we drop the dependence on the initial position $X_0$, it will be self-understood), which satisfies the unperturbed equation $\delta S_{sys}/\delta X(t)[X(t)]\approx 0$. The standard Gaussian integration around $X_{cl}(t)$ gives \cite{FeynmanHibbs}:
\begin{eqnarray}
&&K_t\approx \text e ^{\frac{i}{\hbar}S_{sys}[X_{cl}(t)]}\times\label{K2}\\
&&\int \mathscr D x(t) \text e ^{\frac{i}{\hbar}S_{env}[x(t)]}\text e^{\frac{i}{\hbar}S_{int}[X_{cl}(t),x(t)]} D_t(X_0,X;x(t)),\nonumber
\end{eqnarray}
where $ D_t(X_0,X;x(t))$ is the van Vleck determinant \cite{Schulman2012} for $S_{sys}+S_{int}$. It depends on $x(t)$
through $\delta^2 S_{int}/\delta X(t)\delta X(t')$. This is a quantum leftover of the $E \to S$  back-reaction, which we also neglect, assuming:
\be\label{nbr2}
\left|\frac{\delta^2 S_{sys}}{\delta X(t)\delta X(t')}[X_{cl}(t)]\right|\gg \left|\frac{\delta^2 S_{int}}{\delta X(t)\delta X(t')}[X_{cl}(t),x(t)]\right|,
\ee
which e.g. holds trivially for linearly coupled systems, when $\delta^2 S_{int}/\delta X(t)\delta X(t')=0$.
Then $D_t(X_0,X;x(t))$  reduces to the van Vleck propagator  for $S$ alone, $D_t(X_0,X)$ and can be pulled out of the integral in \eqref{K2}.

\begin{eqnarray}
&&K_t\approx \text e ^{\frac{i}{\hbar}S_{sys}[X_{cl}(t)]}D_t(X_0,X)\times\label{K3}\\
&&\int \mathscr D x(t) \text e ^{\frac{i}{\hbar}S_{env}[x(t)]}\text e^{\frac{i}{\hbar}S_{int}[X_{cl}(t),x(t)]},\nonumber
\end{eqnarray}
where the first two terms define the semi-classical propagator for the central system $K^{sc}_t(X,X_0) \equiv e ^{\frac{i}{\hbar}S_{sys}[X_{cl}(t)]} D_t(X_0,X)$.
The remaining path integral can be represented using the standard operator formalism:
\begin{eqnarray}
\int \mathscr D x(t) \text e ^{\frac{i}{\hbar}S_{env}[x(t)]}\text e^{\frac{i}{\hbar}S_{int}[X_{cl}(t),x(t)]}\equiv \langle x |\hat U_t[X_{cl}]|x_0\rangle,\label{oppath}
\end{eqnarray}
where $\hat U_t[X_{cl}]$ is the effective unitary evolution of the environment with $X_{cl}(t)$ acting as a classical force:
\be\label{Ut}
i\hbar \frac{d\hat U_t}{dt}=\left(\hat H_{env}+\hat H_{int}[X_{cl}(t)]\right)\hat U_t,
\ee
where $\hat H_{env}$, $\hat H_{int}$ are the Hamiltonians corresponding to the actions $S_{env}, S_{int}$ respectively. In what follows we use hats to denote operators.
Thus from (\ref{K2}, \ref{nbr2}, \ref{oppath}, \ref{Ut}) we obtain the propagator in the recoil-less limit:
\be
K_t(X,X_0;x,x_0)\approx K^{sc}_t(X,X_0) \langle x |\hat U_t[X_{cl}]|x_0\rangle. \label{Krecoil}
\ee
Assuming a product initial state $\varrho(0)=\varrho_{0S}\otimes\varrho_{0E}$, we can use  (\ref{Krecoil}) to construct the approximate solution for the full system-environment state $\varrho_{S:E}$. We obtain it in the form of
partial matrix elements between the position states of the central system:
\ben
&&\langle X'|\varrho_{S:E}(t)|X\rangle \approx\nonumber\\
 &&\int\int dX_0 dX'_0 \langle X'_0|\varrho_{0S}| X_0\rangle K^{sc}_t(X,X_0)^*K^{sc}_t(X',X'_0)\nonumber\\
&& \times \hat U_t[X_{cl}]\varrho_{0E}\hat U_t[X'_{cl}]^\dagger.\label{preglowny}
\een
We next specify to the most common situation when the environment is composed out of a number of subenvironments or modes, denoted $E_k$, e.g. a collection of harmonic oscillators. Furthermore, we assume there are not direct interactions between the parts of the environment, only the central interactions so that $\hat H_{env}=\sum_k \hat H_k$ and $\hat H_{int} = \sum_k \hat H_{int}^k$,
where $k$ labels the subenvironments, $\hat H_k$,  $\hat H_{int}^k$  are the self-energy and interaction Hamiltonians of the $k$-th subenvironment respectively.  It is immediate to see that due to that, the effective evolution of the environment 
has  a product structure  $\hat U_t[X_{cl}]=\bigotimes_k \hat U^{k}_t[X_{cl}]$, where each $\hat U^{k}_t[X_{cl}]$ satisfies the corresponding equation \eqref{Ut}. Substituting it into \eqref{preglowny}  and tracing out subenvironments assumed to be unobserved, $E_{uno}$, we obtain the desired solution for the partially traced state \eqref{rseobs}:
\begin{eqnarray}
&&\langle X'|\varrho_{S:E_{obs}}(t)|X\rangle \approx \nonumber\\
&&\int\int dX_0 dX'_0 \langle X'_0|\varrho_{0S}| X_0\rangle K^{sc}_t(X,X_0)^*K^{sc}_t(X',X'_0)\nonumber\\
&&\times  F[X_{cl}(t),X'_{cl}(t)] \bigotimes_{k\in E_{obs}}  \hat U^{k}_t[X_{cl}]\varrho_{0k}\hat U^{k}_t[X'_{cl}]^\dagger.\label{glowny}
\end{eqnarray}
Here we assumed the usual product initial state $\varrho_{0E} =\bigotimes_k \varrho_{0k}$, where $\varrho_{0k}$ is the initial state of the $k$-th subenvironment. Moreover, $X_{cl}(0)=X_0$, $X_{cl}(t)=X$, $X'_{cl}(0)=X'_0$, $X'_{cl}(t)=X'$ and:
\be
F[X_{cl}(t),X'_{cl}(t)]\equiv\prod_{k\in E_{uno}}\tr\left(\hat U^{k}_t[X'_{cl}]^\dagger\hat U^{k}_t[X_{cl}]\varrho_{0k}\right)
\ee
is the influence functional \cite{Feynman1963, FeynmanHibbs}. This is a general, hybrid solution with effectively classical central system, driving quantum environment. The, admittedly coarse, approximation \eqref{glowny} is enough for our purposes.

In what follows we will specify to one of the paradigmatic models of open quantum systems, linear Quantum Brownian Motion  model  (see e.g. \cite{Ullersma1966, Hu1992, Haake1993}), as an example.  
It is described by Lagrangeans: $L_{sys}=1/2(M \dot X ^2-M\Omega ^2 X^2)$, $L_{env}=\sum_k1/2 (m_k \dot x_k^2-m_k\omega_k ^2 x_k^2)$, $L_{int}=-X\sum_k C_k x_k$ and the corresponding actions. 
It is easy to see that the no-recoil condition \eqref{nbr} will be satisfied when: 
\be\label{norecoil}
 \frac{C_k}{M\Omega^2}\ll 1,
\ee
so that  $X_{cl}(\tau)$ are the ordinary oscillator trajectories, and the condition \eqref{nbr2} is trivial due to the linearity 
in $X$ of the interaction term. 


The influence functional for QBM has been first calculated in \cite{Feynman1963} 
for the physically most relevant situation of the thermal environments and has the well-known form\cite{Feynman1963, Caldeira1983,Breuer2002, Joos2003}:

\begin{eqnarray}
&&F[X_{cl}(t),X'_{cl}(t)]=\nonumber\\
&&\text{exp}\left\{-\frac{1}{\hbar}\int_0^t d\tau\int_0^\tau d\tau' \Delta(\tau)\nu(\tau-\tau')\Delta(\tau')\right\}\label{ReF}\\
&&\times\text{exp}\left\{-\frac{i}{\hbar}\int_0^t d\tau\int_0^\tau d\tau' \Delta(\tau)\eta(\tau-\tau')\bar X_{cl}(\tau')\right\},\label{ImF} 
\end{eqnarray}
where $\Delta(\tau)\equiv X_{cl}(\tau)-X'_{cl}(\tau)$ is the trajectory difference, $\bar X_{cl}(\tau)\equiv (1/2)( X_{cl}(\tau)+X'_{cl}(\tau))$ is the trajectory average,
and $\nu(\tau), \eta(\tau)$ are the noise and dissipation kernels respectively \cite{Feynman1963, Caldeira1983,Breuer2002, Joos2003, Schlosshauer2007}:
\begin{eqnarray}
&&\nu(\tau)\equiv\int d\omega J_{uno}(\omega) \mathrm{cth}\left(\frac{\hbar\omega\beta}{2}\right)\cos\omega\tau, \label{nk}\\
&&\eta(\tau)\equiv \int d\omega J_{uno}(\omega) \sin\omega\tau,
\end{eqnarray}
with $\beta=1/(k_BT)$ denoting the inverse temperature and 
$J_{uno}(\omega)\equiv\sum_{k\in E_{uno}} C_k^2/(2 m_k\omega_k)\delta(\omega-\omega_k)$ is the spectral density
of the unobserved part of the environment. The modulus of $F[X_{cl}(t),X'_{cl}(t)]$ controls the decoherence process.

\section{Distinguishability of local states and Quantum Fisher Information kernel} 
To understand what information about $S$ is available locally in the the environment, we need the local states of each subenvironment $E_k$, 
as these are the states that fully determine results of local measurements for each of the observers. 
$\varrho_{k}(t)=\tr_{E_1\dots  \cancel{E_k} \dots}\tr_S \varrho_{S:E_{obs}}$.
The detailed calculation, relying on (\ref{norecoil}), is presented in the Appendix ~\ref{partialtrace}. The result is: 
$\varrho_{k}(t) \approx \int dX_0 p(X_0) \varrho^k_t[X^0_{cl}],$ and similarly for the whole observed fraction of the environment $E_{obs}$:
\ben
\varrho_{E_{obs}}(t) \approx
 \int dX_0 p(X_0) \bigotimes_{k\in E_{obs}} \varrho^k_t[X^0_{cl}],\label{rk}
\een 
where:
\be
\varrho^k_t[X_{cl}]\equiv \hat U^{k}_t[X_{cl}]\varrho_{0k}\hat U^{k}_t[X_{cl}]^\dagger \label{rt},
\ee 
are conditional states of $E_k$, $p(X_0)\equiv \langle X_0|\varrho_{0S}|X_0\rangle$, and $X^0_{cl}$ is the classical trajectory with the endpoint $0$: $X^0_{cl}(0)=X_0,\ X^0_{cl}(t)=0$. 
In the case of linear QBM, the evolution law (\ref{Ut}), satisfied by  $\hat U^{k}_t[X_{cl}]$, 
describes a harmonic oscillator forced along the classical trajectory $X_{cl}$.
It has a well known solution, which in the interaction picture reads (we present it in Appendix~\ref{app:overlap} for completeness):
\be 
\hat U^{k}_t[X_{cl}]=e^{i\zeta_k(t)}\hat D\left(-\frac{iC_k}{\sqrt{2\hbar m_k\omega_k}}\int_0^t
d\tau e^{i\omega_k \tau} X_{cl}(\tau)\right),\label{D}
\ee
where $\zeta_k(t)$ is an irrelevant phase factor and $\hat D(\alpha)\equiv\exp(\alpha \hat a^\dagger-\alpha^*\hat a)$ is the standard optical displacement operator.
The local states of $E_k$ are mixtures of oscillator states (\ref{rt}), forced along  $X^0_{cl}$. They are parametrized  by the central system's initial position $X_0$ (cf. \cite{Tuziemski2015}), spread with the probability $p(X_0)$.

The information content of the fragment $E_k$ is determined by the distinguishability  of  the local states $\varrho^k_t[X_{cl}]$ for different trajectories. We will consider a general $X_{cl}(\tau)$, which can later be specified to $X^0_{cl}(\tau)$. 
Among the available distinguishability measures \cite{Fuchs1999}, a particularly convenient one is the generalized overlap  $B(\varrho,\sigma)\equiv \tr\sqrt{\sqrt\varrho\sigma\sqrt\varrho}$. It provides both lower 
and upper bounds for such operational quantities as the probability of error and to the quantum Chernoff information \cite{Calsamiglia2008} and is a good compromise between computability and a 
clear operational meaning. We define the generalized overlap for the conditional states of the $k$-th environment:
\be\label{Bk}
B_k[\Delta, t]\equiv B(\varrho^k_t[X_{cl}],\varrho^k_t[X'_{cl}])
\ee
(from the definition and \eqref{rt},\eqref{D}, $B$ depends only on the difference of trajectories $\Delta$). 
For thermal $\varrho_{0k}$, the overlap \eqref{Bk} was found in \cite{Tuziemski2015} (see Appendix~\ref{app:overlap}):
\begin{eqnarray}
&&B_{k}[\Delta, t]=\label{B1}\\
&&\mathrm{exp}\left\{-\frac{C_k^2}{4\hbar m_k\omega_k}\mathrm{th}\left(\frac{\hbar\omega_k\beta}{2}\right)\left|\int_0^t d\tau\mathrm e^{i\omega_k\tau}\Delta(\tau)\right|^2\right\}.\nonumber
\end{eqnarray}

A single environmental mode typically carries a vanishingly small amount of useful information.    
To decrease the discrimination error, it is beneficial to combine the modes into groups, called macrofractions \cite{Korbicz2014},  scaling with the total number of observed modes.
In our case, we consider the whole observed environment $E_{obs}$. Since there are no direct interactions in the environment, the conditional states of the observed fraction  are  
products, cf. \eqref{rk}: $\varrho^{obs}_t[X_{cl}]\equiv \bigotimes_{k\in E_{obs}} \varrho^k_t[X_{cl}]$. The generalized overlap factorizes w.r.t. the tensor product, so there is no
quantum metrological advantage here \cite{Giovannetti2006}, but still there is a classical one \cite{Korbicz2014} :
\begin{eqnarray}
&&B_{obs}[\Delta, t]\equiv  B(\varrho^{obs}_t[X_{cl}],\varrho^{obs}_t[X'_{cl}])=\prod_{k\in E_{obs}}B_{k}[\Delta, t]\nonumber\\
&&=\mathrm{exp}\left\{-\sum_{k\in E_{obs}}\frac{C_k^2}{4\hbar m_k\omega_k}\mathrm{th}\left(\frac{\hbar\omega_k\beta}{2}\right)\left|\int_0^t d\tau\mathrm e^{i\omega_k\tau}\Delta(\tau)\right|^2\right\}\nonumber\\
&&\equiv\mathrm{exp}\left\{-\frac{1}{\hbar}\int_0^t d\tau\int_0^\tau d\tau' \Delta(\tau)\phi(\tau-\tau')\Delta(\tau')\right\}, \label{Bmac}
\end{eqnarray}
where  in the last step we passed to the continuum limit and introduced a  new kernel:
\begin{eqnarray}
\phi(\tau)\equiv\int_0^\infty d\omega J_{obs}(\omega) \mathrm{th}\left(\frac{\hbar\omega\beta}{2}\right)\cos\omega\tau\label{qfik},
\end{eqnarray}
called quantum Fisher information kernel. Here $J_{obs}(\omega)\equiv\sum_{k\in E_{obs}} C_k^2/(2 m_k\omega_k)\delta(\omega-\omega_k)$ is the spectral density corresponding to the observed environment.
Note that, quite surprisingly, the  QFI kernel and the overlap \eqref{Bmac} have almost identical structure to the noise kernel \eqref{nk} and the real part of the influence functional \cite{Feynman1963, Caldeira1983,Breuer2002, Joos2003}, 
the only difference being in the reversed temperature dependence \cite{Tuziemski2015,  Tuziemski2016}. It can be intuitively  understood by recalling that here the higher the
temperature the more efficient the decoherence but also more noisy the environment.  The name QFI kernel comes from the observation that the QFI of the $X_0$ phase imprinting: 
$\varrho_k(X)\equiv \mathrm e^{-i/\hbar X_0C_k\hat x_k}\varrho_{0k}\mathrm e^{i/\hbar X_0C_k\hat x_k}$
is proportional to the integrand of \eqref{qfik} (see e.g. \cite{Toth2014}). 

We want a fair comparison of the decohering power and the information content of the observed environment, 
so we assume equal spectral densities for the unobserved and the observed fractions $J_{obs} = J_{uno}\equiv J(\omega)$ and choose it to be in the Lorenz-Drude form \cite{Breuer2002, Joos2003, Schlosshauer2007}: 
\be
J(\omega)=\frac{2M\gamma}{\pi}\omega \frac{\Lambda^2}{\Lambda^2+\omega^2}, 
\ee
where $\Lambda$ is the cutoff frequency and $\gamma$ is the effective coupling strength.

\section{The information gap} 
For our demonstration it is enough to use the highly popular Caldeira-Leggett limit \cite{Caldeira1983, Schlosshauer2007},
$k_BT/\hbar\gg\Lambda\gg\Omega.\label{CL}$, which is the high-temperature, hight-cutoff limit.
The behavior of the influence functional in this limit is emblematic to the whole decoherence theory and can be obtained e.g. by approximating $\text{cth} x \approx x^{-1}$ in the 
noise kernel (\ref{nk}) and then using $\Lambda e^{-\Lambda\tau} \approx \delta(\tau)$ valid for  $\tau\gg \Lambda^{-1}$ (or using the Matsubara representation \cite{Mahan2000}). 
This leads to the celebrated result  that decoherence becomes efficient at lengths above  the thermal de Broglie wavelength $\lambda_{dB}^2 = \hbar^2/2Mk_BT$  
\cite{Papa1968, Caldeira1983, Moore1986, Joos2003}:
\be\label{F approx}
\Big|F[X_{cl}(t),X'_{cl}(t)]\Big|\approx\mathrm{exp}\left[- \frac{\gamma}{\lambda_{dB}^2} \int_0^t d\tau\Delta(\tau)^2\right],
\ee                                                                                                                                                                                                                                                                                                                                                                                                                                                                                                                                                                                  
and for times larger than  the decoherence time \cite{Moore1986} $t_{dec}= 1/\gamma(\lambda_{dB}/d)^2$, where $d$ is a given separation and $\gamma^{-1}$  is related to the relaxation time.

The QFI kernel can be studied in the similar way, approximating $\text{th} x \approx x$ in \eqref{qfik} and passing to a large $\Lambda\tau$. 

\begin{align}
& \phi(\tau) \underset{\hbar\Lambda\beta\ll 1}{\approx} \frac{\gamma M\hbar \beta \Lambda^2}{\pi}\int_0^\infty d\omega \frac{\omega^2}{\omega^2+\Lambda^2}\cos(\omega\tau)\\
& =\frac{\gamma M\hbar \beta \Lambda^2}{\pi}\left(\int_0^\infty d\omega \cos(\omega\tau) - \Lambda^2\int_0^\infty d\omega\frac{\cos(\omega\tau)}{\omega^2+\Lambda^2}\right)\\
& = \gamma M \hbar \beta  \left(\Lambda^2\delta(\tau)-\frac{1}{2}\Lambda^3 e^{-\Lambda|\tau|}\right). \label{phiapprox}
\end{align}
We define $f(\tau ) \equiv \Lambda e^{-\Lambda\tau }$ for $\tau>0$ so that $\Lambda^3 e^{-\Lambda(\tau-\tau')} = d^2/d\tau'^2f(\tau-\tau')$.
We can then calculate the integral with the second term of \eqref{phiapprox} integrating by parts two times:
\begin{align}
& \int_0^\tau d\tau'\Lambda^3 e^{-\Lambda(\tau-\tau')} \Delta(\tau')=\int_0^\tau d\tau'\ddot f(\tau-\tau')\Delta(\tau')\\
& =\dot f(0_+) \Delta(\tau)  -\dot f(\tau) \Delta (0) - f(0)\dot\Delta(\tau) + f(\tau)\dot\Delta(0)\\
& + \int_0^\tau d\tau' f(\tau-\tau')\ddot \Delta(\tau'),
\end{align}
here the dot means $d/d\tau'$, so that  $\dot f(0_+)=+\Lambda^2$. In the large cut-off limit it is justified to assume $\tau \gg \Lambda^{-1}$, i.e. we consider timescale much larger than the one set by the cut-off, similarly as it is done analyzing the influence functional \cite{Caldeira1983}. Then the boundary terms containing $f(\tau)$ and $\dot f(\tau)$ can be neglected and moreover we can substitute $f(\tau) \approx \delta(\tau)$ in the last integral to obtain:
\begin{align}
& \int_0^\tau d\tau'\Lambda^3 e^{-\Lambda|\tau-\tau'|} \Delta(\tau') \underset{\Lambda\tau\gg 1}{\approx} \Lambda^2 \Delta(\tau) - \Lambda \dot\Delta(\tau) + \ddot \Delta(\tau)\\
& =  \Lambda^2 \Delta(\tau) -\Lambda\Omega \Delta\left(\tau+\frac{\pi}{2\Omega}\right) -\Omega^2 \Delta(\tau)\\
& \underset{\Lambda \gg \Omega}{\approx} \Lambda^2 \Delta(\tau), \label{L2D}
\end{align} 
where we used the fact that $\Delta(\tau)$ is a difference of two oscillator trajectories, so that it satisfies: $\dot\Delta(\tau) = \Omega\Delta(\tau+\pi/2\Omega)$ and $\ddot \Delta(\tau)=-\Omega^2\Delta(\tau)$. Finally, we neglected the $\Omega$-terms because $\Lambda \gg \Omega$. More generally, $\dot \Delta(\tau)$, $\ddot\Delta(\tau)$ are inverse proportional to the system's evolution time-scale, which is the slowest time-scale, and hence those terms can be omitted compared to the $\Lambda^2$ term. Using \eqref{L2D} and \eqref{phiapprox} we obtain our main result -- the  generalized overlap between the environmental macro-states. It is both remarkably simple and similar to \eqref{F approx}:
\begin{align}
B[\Delta, t] \approx \mathrm{exp}\left[- \frac{\gamma}{\lambda^2_{dist}}\int_0^t d\tau\Delta(\tau)^2\right].\label{Bfinal}
\end{align}
The expression \eqref{Bfinal} immediately implies that the distinguishability process is described by its own lengthscale, which we call the distinguishability length:
\be
\lambda_{dist}^2\equiv \frac{2k_B T}{M\Lambda^2}=\frac{\hbar}{M\Lambda}\left(\frac{2k_BT}{\hbar\Lambda}\right)\label{ldist}
\ee 
and happens on the associated distinguishability timescale:
\be\label{tdist}
t_{dist} = \frac{1}{\gamma}\left(\frac{\lambda_{dist}}{d}\right)^2.
\ee
Surprisingly, \eqref{ldist} does not depend on $\hbar$ in the leading order. It is the lengthscale at which the energy of the "cut-off oscillator" of mass $M$ and frequency $\Lambda$ equals the thermal energy:
 $M\Lambda^2\lambda_{dist}^2/2=k_BT$. The cut-off dependence of \eqref{ldist} can be understood  recalling that the cut-off defines the shortest lengthscale in the environment. Indeed, 
 \eqref{ldist} can be expressed as the characteristic length of the cut-off oscillator ,
 $\sqrt{\hbar /M\Lambda}$, rescaled by the ratio of the thermal energy to the cut off energy, $\sqrt{2k_BT/\hbar\Lambda}$.
Of course the higher order terms in the $\text{th}(\hbar\beta\omega/2)$ expansion in \eqref{Bmac},\eqref{qfik} will contribute $O(\hbar^2)$ terms to \eqref{ldist}. There is clearly a competition in \eqref{ldist} between the temperature $T$, which degrades the discriminating ability of the environment  and the cut-off frequency $\Lambda$ which increases it. The relative difference between the two  lengthscales:
\be
\frac{\lambda_{dist} - \lambda_{dB}}{\lambda_{dB}} \approx 2\frac{k_BT}{\hbar\Lambda}\gg 1,
\ee
shows that there is a ``resolution gap'' between the decoherence and the distinguishability accuracy \cite{remark}. The environment decoheres the system at shorter lengthscales than those at which information can be extracted from it, i.e. a part of 
 information stays bounded in the environment. The timescales are separated even stronger: 
\be
\frac{t_{dist}}{t_{dec}}=4\left(\frac{k_BT}{\hbar\Lambda}\right)^2,
\ee 
meaning the distinguishability process takes much longer time than the decoherence for the same separation. 
This is in accord with the earlier results for generic finite dimensional systems; see e.g. \cite{Korbicz2017}. For molecular environments $\Lambda \sim 10^{13}$Hz and  at $T\sim 300$K, $\lambda_{dist}/\lambda_{dec} \sim 10$, 
$t_{dist}/t_{dec} \sim 100$, which is still orders of magnitude shorter for macroscopic bodies than typical relaxation times \cite{Moore1986}.

As a by-product we obtain a type of  information gain vs. disturbance relation (see e.g. \cite{Buscemi2008}), where the disturbance is represented by the decoherence efficiency:
\be
\lambda_{dB}\lambda_{dist}= \frac{\hbar}{ M\Lambda}.
\ee
The right hand side does not depend on the state of the environment (encoded in the temperature) and is the square of the characteristic length of the cut-off oscillator. More generally,
passing to the Fourier transforms of the noise, dissipation, and QFI kernels, denoted by the tilde, we obtain the following relations, true for thermal environments:
\ben
&&\tilde\phi(\omega)=\tilde\nu(\omega)\text{th}^2\left(\frac{\hbar\omega\beta}{2}\right),\\
&&\tilde\phi(\omega)=i\tilde\eta(\omega)\text{th}\left(\frac{\hbar\omega\beta}{2}\right).
\een
They resemble the celebrated fluctuation-dissipation relation \cite{Kubo1966, Hu1992, Campisi2011}:
\be
\tilde\nu(\omega)=i\tilde\eta(\omega)\text{cth}\left(\frac{\hbar\omega\beta}{2}\right),
\ee
but connect dissipation and noise to information accumulation in the environment. 
These interesting relations will be investigated further elsewhere.

\section{Conclusions} 
We have shown here, using the celebrated model of Caldeira and Leggett as an example, that there is an information gap between what environment learns, decohering the system, and what can be extracted from it via measurements, i.e. some information stays bounded in the environment. For that, we have developed a series of rather non-trivial and non-standard tools, including a path integral recoiless limit, a hybrid quantum-classical extended state solution \eqref{glowny}, and Quantum Fisher information kernel. Our results uncover the existence of a new lengthscale, determining information content of the environment and complementary to the celebrated thermal de Broglie decoherence lengthscale.  

The unorthodox point of view taken here, i.e. that of the environment instead of the the central system, has been motivated by the modern developments of the decoherence theory \cite{Zurek2009, Korbicz2014, Brandao2015}, explaining the apparent objectivity of the macroscopic world through redundantly stored information in the environment.  From this perspective, the solution \eqref{glowny} can approximate an SBS state, storing an objective position of the central system, in the semi-classical approximation. We hypothesize that approach to objectivity is possible only in such a limit, when the central system is macroscopic enough, making objectivity a macroscopic phenomenon. 

We acknowledge the support by Polish National Science Center (NCN) (Grant No. 2019/35/B/ST2/01896). JKK acknowledges discussions with J. Tuziemski in the early stages of the work.

\appendix

\section{Alternative derivation of the hybrid solution }\label{alternative}
In the particular case of  a linear QBM model, the hybrid $SE_{obs}$ solution from the main text:
\begin{eqnarray}
&&\langle X'|\varrho_{S:E_{obs}}(t)|X\rangle \approx \nonumber\\
&&\int\int dX_0 dX'_0 \langle X'_0|\varrho_{0S}| X_0\rangle K^{sc}_t(X,X_0)^*K^{sc}_t(X',X'_0)\nonumber\\
&&\times  F[X_{cl}(t),X'_{cl}(t)] \bigotimes_{k\in E_{obs}}  \hat U^{k}_t[X_{cl}]\varrho_{0k}\hat U^{k}_t[X'_{cl}]^\dagger.\label{app:glowny}
\end{eqnarray}
can be also obtained in the following way: Forgetting for a moment the evolution of the environment, the central system is  a forced harmonic oscillator with a well-known solution for the propagator \cite{Feynman1963}. It is determined by the action: 
\ben
&& S[x(t)]=\frac{M\Omega}{2\sin\Omega t}\big[(X^2+X_0^2)\cos\Omega t- 2X X_0\big]+\nonumber\\
&& \frac{1}{\sin\Omega t} \sum_k C_k\int_0^t d\tau \left[X\sin\Omega \tau + X_0 \sin\Omega (t-\tau)\right] x_k(\tau) \nonumber\\
&& -\frac{\Omega}{\sin\Omega t}\sum_{k,l} \frac{C_k C_l}{M\Omega^2} \int_0^t d\tau\int_0^{\tau} d\tau' \sin\Omega \tau' \sin\Omega(t-\tau)\nonumber\\
&&\times x_k(\tau)x_l(\tau').
\een
Neglecting the last term, using the resulting action to construct the propagator for the global state, and changing to the operator picture for the environmental degrees of freedom, we obtain the solution (\ref{glowny}).

\section{Tracing over the central system}\label{partialtrace}
Here we calculate the trace over the central system $S$ of the hybrid solution \eqref{glowny}.
We first assume for simplicity only one observed environment and one unobserved. Generalization to multiple environments in both groups will be obvious and we present it at the end. The main idea is to rewrite the trace using the no-recoil condition 
\be\label{app:norecoil}
 \frac{C_k}{M\Omega^2}\ll 1.
\ee
First, we take matrix elements w.r.t. the environment and comeback from the operator form of the environment part of (\ref{glowny}) to the path integral one using:
\be
\int \mathscr D x(t) \text e ^{\frac{i}{\hbar}S_{env}[x(t)]}\text e^{\frac{i}{\hbar}S_{int}[X_{cl}(t),x(t)]}\equiv \langle x |\hat U_t[X_{cl}]|x_0\rangle.\label{app:oppath}
\ee
This gives:
\ben
&&\int dX \langle X;x'| \varrho_{S:E_{obs}}| X;x\rangle =\nonumber\\
&&\int dX_0dX_0'dx_0dx'_0 \langle X'_0|\varrho_{0S}| X_0\rangle \langle x'_0|\varrho_{0E}| x_0\rangle \\
&&\times \int dX  K^{sc}_t(X,X_0)^*K^{sc}_t(X,X'_0)F[X_{cl},X'_{cl}]\label{TrS2}\\
&&\times \int \mathscr D x \mathscr D x' \exp\frac{i}{\hbar}\Big(S_{env}[x] - S_{env}[x'] + \nonumber\\
&&S_{int}[X_{cl},x]- S_{int}[X'_{cl},x']\Big)\label{TrS3}
\een
Because of the tracing, the classical trajectories have the same endpoints $X_{cl}(t)=X_{cl}'(t)=X$, and $x(0)=x_0$, $x(t)=x$ and similarly for $x'(\tau)$. Let us analyze the above expression term by term. It is well known that the semiclassical propagator $K^{sc}_t(X,X_0)$ for harmonic oscillator is equal to the full quantum one. We thus have:
\ben
&&K^{sc}_t(X,X_0)^*K^{sc}_t(X,X'_0)=\\
&&\frac{M\Omega}{2\pi\hbar |\sin\Omega t|} e^{\frac{iM\Omega}{2\hbar\sin\Omega t}\big[(X_0'^2-X_0^2)\cos\Omega t+ 2X \Delta X_0\big]}.\label{KK}
\een
There are $X$-dependent and $X$-independent parts.

The influence functional may be written using path integrals as \cite{FeynmanHibbs}:
\ben
&&F[X_{cl},X'_{cl}]=\int d\tilde y dy_0dy'_0\int \mathscr D y \mathscr D y'\langle y'_0|\tilde\varrho_{0E}| y_0\rangle\nonumber\\
&&\times e^{\frac{i}{\hbar}\big(S_{env}[y] - S_{env}[y'] + S_{int}[X_{cl},y]- S_{int}[X'_{cl},y']\big)},\label{Fpath}
\een
where $\tilde\varrho_{0E}$ is the initial state of the unobserved part of the environment $E_{uno}$, which can be different from the initial state of the observed part $E_{obs}$, $\varrho_{0E}$ in (\ref{TrS2}). The boundary conditions are $y(0)=y_0$, $y'(0)=y'_0$, $y(t)=y'(t)=\tilde y$. The generalization to multiple unobserved environments is straightforward -- the combined influence functional will be a product over $j\in E_{uno}$ of the terms (\ref{Fpath}) for each mode $j$ with $\varrho_{0j}$ initial state, $F=\prod_{j\in E_{uno}} F_j$

The terms of the form $S_{env}[x] - S_{env}[x']$, appearing both in (\ref{TrS3}) and (\ref{Fpath}), do not depend on the integration variable $X$ and thus can pulled in from of the integral over $X$. The interaction terms $S_{int}[X_{cl},y]- S_{int}[X'_{cl},y']$
from \eqref{TrS3}, \eqref{Fpath} will have both $X$-dependent and $X$-independent parts. To separate them, let us parametrize the classical trajectory $X_{cl}$ satisfying the apropriate boundary conditions $X_{cl}(0)=X_0$, $X_{cl}(t)=X$, as below:
\ben
&&X_{cl}(\tau)=X_0 \cos\Omega\tau\left[1-\frac{\sin\Omega \tau}{\sin\Omega t}\right]+X\frac{\sin\Omega\tau}{\sin\Omega t}\\
&&\equiv X_0a(\tau)+X\frac{\sin\Omega\tau}{\sin\Omega t}.\label{Xcl'}
\een
Then it is easy to see that:
\ben
&&S_{int}[X_{cl},x]- S_{int}[X'_{cl},x']=\nonumber\\
&&-C\int_0^t d\tau a(\tau) \left[X_0x(\tau)-X_0'x'(\tau)\right]\label{S-S}\\
&& -\frac{C X}{\sin\Omega t}\int_0^t d\tau \sin\Omega\tau \left[x(\tau)-x'(\tau)\right]
\een
We now combine the $X$-dependent factors from \eqref{TrS3},\eqref{KK},\eqref{Fpath} and integrate them over $X$:
\begin{widetext}
\ben
&&\frac{M\Omega}{2\pi\hbar |\sin\Omega t|}\int dX \exp\Bigg\{\frac{i X}{\hbar\sin\Omega t}\bigg[M\Omega\Delta X_0-C\int_0^t d\tau \sin\Omega\tau \left[x(\tau)-x'(\tau)+y(\tau)-y'(\tau)\right]\bigg]\Bigg\}\\
&&=\frac{M\Omega}{2\pi\hbar |\sin\Omega t|}\int dX \exp\Bigg\{\frac{i X M\Omega^2}{\hbar\sin\Omega t}\bigg[\frac{\Delta X_0}{\Omega}-\frac{C}{M\Omega^2}\int_0^t d\tau \sin\Omega\tau \left[x(\tau)-x'(\tau)+y(\tau)-y'(\tau)\right]\bigg]\Bigg\}\\
&&\approx \frac{M\Omega}{2\pi\hbar |\sin\Omega t|}\int dX \exp\bigg(\frac{i X M\Omega\Delta X_0}{\hbar\sin\Omega t}\bigg) =\delta\left(\Delta X_0\right),\label{delta}
\een
\end{widetext}
where in the crucial step we used the recoilless condition (\ref{norecoil}) and neglected the action integral. We can now comeback to the main integral (\ref{TrS2},\ref{TrS3}). The delta function (\ref{delta}) forces the trajectories $X_{cl}(\tau)$ and $X'_{cl}(\tau)$ to be equal as it forces $X_0=X_0'$ (the endpoints are the same in this calculation as we are calculating the trace over $X$). This immediately forces the influence functional (\ref{Fpath}) to be equal to one since:
\ben
&&F[X_{cl},X_{cl}]=\tr\left(\hat U_t[X_{cl}]\tilde\varrho_{0E}\hat U_t[X_{cl}]^\dagger\right)=1
\een
The $X$-independent part of (\ref{KK}) will be equal to one too, since $X_0'^2-X^2_0=0$. We are thus left with the following integral:
\begin{widetext}
\ben
&&\int dX \langle X;x'| \varrho_{S:E_{obs}}| X;x\rangle =\int dX_0 p(X_0)\int dx_0dx'_0 \langle x'_0|\varrho_{0E}| x_0\rangle \\
&&\times  \int \mathscr D x \mathscr D x'\exp\frac{i}{\hbar}\Big(S_{env}[x] - S_{env}[x'] -C\int_0^t d\tau a(\tau) \left[X_0x(\tau)-X_0'x'(\tau) \right]\Big)\label{TrS4}\\
&&=\int dX_0 p(X_0)\int dx_0dx'_0 \langle x'_0|\varrho_{0E}| x_0\rangle \int \mathscr D x \mathscr D x' e^{\frac{i}{\hbar}\big(S_{env}[x] +S_{int}[X^0_{cl},x] - S_{env}[x']-S_{int}[X^0_{cl},x']\big)}\label{TrS5}\\
&&= \int dX_0 p(X_0) \langle x'|U_t[X^0_{cl}]\varrho_{0E}U_t[X^0_{cl}]^\dagger|x\rangle\label{TrS6},
\een
\end{widetext}
where in (\ref{TrS5}) we came back to the operator picture using \eqref{app:oppath} and introduced:
\be
p(X_0)\equiv \langle X_0|\varrho_{0S}|X_0\rangle,
\ee
which is the initial distribution of the central system's position.  Above, $X^0_{cl}$ is the classical trajectory with the endpoint $0$:
\be 
X^0_{cl}(0)=X_0,\ X^0_{cl}(t)=0. 
\ee
It appears by comparing the action integral in the exponent of (\ref{TrS4}) to (\ref{Xcl'}) with $X=0$. Having the result (\ref{TrS6}) for a single degree of freedom of the observed environment, we can now apply it to the initial task with multiple environments. Performing the above calculations for each degree of freedom $j$ we finally obtain:
\ben
&&\varrho_{k}(t)=\tr_{E_1\dots \cancel{E_k}\dots}
\int dX \langle X| \varrho_{S:E_{obs}}| X\rangle \approx\\
&&\int dX_0 p(X_0) \tr_{E_1\dots \cancel{E_k}\dots}\bigotimes_{j\in E_{obs}}  \hat U^{j}_t[X^0_{cl}]\varrho_{0j}\hat U^{j}_t[X^0_{cl}]^\dagger\nonumber\\
&& =\int dX_0 p(X_0) \hat U^{k}_t[X^0_{cl}]\varrho_{0k}\hat U^{k}_t[X^0_{cl}]^\dagger \\
&&\equiv \int dX_0 p(X_0) \varrho^k_t[X^0_{cl}],
\een 
where the approximation signalizes that we have used the no-recoil condition (\ref{norecoil}) and we have defined 
\be
\varrho^k_t[X_{cl}]\equiv \hat U^{k}_t[X_{cl}]\varrho_{0k}\hat U^{k}_t[X_{cl}]^\dagger. 
\ee

\section{Generalized overlap for thermal QBM}\label{app:overlap}

For completeness' sake, we present here the derivation of the generalize overlap \eqref{Bk}  from \cite{Tuziemski2015}.
We first solve the effective dynamics for the environmental modes, resulting from \eqref{Ut}. In the case of the linear QBM considered here, the effective Hamiltonian $\hat H_{eff} \equiv \hat H_{env} + \hat H_{int}[X_{cl}]$ decomposes of course w.r.t. the subenvironments and for the $k$-th subevironment has a  simple form:
\be
\hat H_{eff}^k = \frac{\hat p_k^2}{2m_k}+\frac{m_k\omega_k^2 \hat x_k^2}{2} - C_k X_{cl}(t) \hat x_k,
\ee
where $\hat x_k, \hat p_k$ are the canonical observables. This is a standard forced harmonic oscillator. It can be solved in many ways, the fastest being by passing to the interaction picture:
\be
 \hat H_{eff}^k (t)= - C_k X_{cl}(t) \hat x_k(t), \label{Heff}
\ee
where $\hat x_k(t)= \sqrt{\hbar/2m_k\omega_k}(e^{-i\omega_kt} \hat a + e^{i\omega_kt} \hat a_k^\dagger)$ with $\hat a_k, \hat a^\dagger_k$ being the corresponding anihilation and creation operators.
Using the fact that \eqref{Heff} commute for different times to a number: $[\hat H_{eff}^k (t),\hat H_{eff}^k (t')]=ic(t,t')$, one can use the Baker–Campbell–Hausdorff formula formula to calculate the evolution
via $\lim_{n\to\infty} \big(\prod_{r=1}^n \exp[-i/\hbar\hat H^k_{eff}(t_r)\Delta t]\big)$, $\Delta t\equiv t/n$, $t_r\equiv r\Delta t$:
\ben
&&\lim_{n\to\infty} \big(\prod_{r=1}^n \exp[-\frac{i}{\hbar}\hat H^k_{eff}(t_r)\Delta t]\big)=\nonumber\\
&&e^{i\zeta(t)}\mathrm{exp}\left(-\frac{i}{\hbar}\int_0^t d\tau  \hat H_{eff}^k (\tau)\right)=\\
&&e^{i\zeta(t)} \mathrm{exp}\left[-i\frac{C_k}{\sqrt{2\hbar m_k\omega_k}}\left(\int_0^td\tau X_{cl}(\tau)e^{i\omega_k\tau} + c.c \right)\right]\nonumber\\
\label{app:Ueff}
\een
where $\zeta(t)$ is some phase factor, that as we will see below will be unimportant for our calculations. Defining:
\be\label{alpha}
\alpha(t)\equiv -\frac{iC_k}{\sqrt{2\hbar m_k\omega_k}}\int_0^td\tau X_{cl}(\tau)e^{i\omega_k\tau},
\ee
the exponent in \eqref{app:Ueff} becomes 
the standard displacement operator $\hat D(\alpha(t))$, so that in the interactrion picture we obtain:
\be
\hat U_t^k[X_{cl}]=e^{i\zeta(t)} \hat D(\alpha(t))\label{app:Uk},
\ee
which is the expression \eqref{Ut}.

We now calculate the single system generalized overlap \eqref{Bk}. 
To simplyfy the notation, we will drop the index  $k$ in all the objects that define \eqref{Bk}, since the calculation is the same for every mode.  Using the definition of the generalized overlap together with the operator identity, following from the spectral theorem $\sqrt{UAU^\dagger}=U\sqrt A U^\dagger$, we obtain:  
\ben
&&B[\Delta, t]= \nonumber\\
&&\tr\sqrt{\sqrt{\varrho_0}\hat U_t[X_{cl}']^\dagger\hat U_t[X_{cl}]\varrho_0\hat U_t[X_{cl}]^\dagger\hat U_t[X_{cl}']\sqrt{\varrho_0}},\label{app:B}
\een
where we have pulled the extreme left and right unitaries out of the both square roots and used the cyclic property of the trace to cancel them out.
From \eqref{app:Uk} we obtain that modulo phase factors, which cancel in \eqref{app:B}:
\be
\hat U_t[X_{cl}']^\dagger\hat U_t[X_{cl}] \simeq \hat D(\alpha(t)-\alpha'(t))\equiv\hat D(\eta_t), \label{UU}
\ee
where we could use the interaction picture expression \eqref{app:Uk} since free evolutions cancel  and we introduced $\eta_t\equiv \alpha(t)-\alpha'(t)$ for a later convenience.
Next, assuming that $\varrho_{0}$ is a thermal state,
we use the well-known coherent state representation for the middle $\varrho_0$ under the square root in \eqref{app:B}:
\be
\varrho_0=\int \frac{d^2\gamma}{\pi\bar n} e^{-|\gamma|^2/\bar n}\ket\gamma\bra\gamma,\label{P-rep}
\ee 
where $\bar n=1/(e^{\hbar\beta\omega}-1)$ is the mean excitation number at the inverse temperature $\beta$.
Denoting the Hermitian operator under the square root in \eqref{app:B} by $\hat A_t$, we obtain:
\ben
&&\hat A_t = \int\frac{d^2\gamma}{\pi\bar n}e^{-|\gamma|^2/\bar n}\sqrt{\varrho_0}\hat D(\eta_t)\ket\gamma\bra\gamma\hat D(\eta_t)^\dagger\sqrt{\varrho_0}\\
&&=\int \frac{d^2\gamma}{\pi\bar n} e^{-|\gamma|^2/\bar n}\sqrt{\varrho_0}\ket{\gamma+\eta_t}\bra{\gamma+\eta_t}\sqrt{\varrho_0}.
\een
To perform the square roots above, we now use the Fock representation of the thermal state:
\be
\varrho_0=\sum_n\frac{\bar{n}^n}{(\bar n +1)^{n+1} }\ket n\bra n, \label{Fock}
\ee
so that:
\begin{eqnarray}
\hat A_t&=&\int\frac{d^2\gamma}{\pi\bar n}e^{-\frac{|\gamma|^2}{\bar n}}\sum_{m,n}\sqrt{\frac{\bar n^{m+n}}{(\bar n +1)^{m+n+2}}}\times\nonumber\\
& \times& \langle n|\gamma+\eta_t\rangle\langle\gamma+\eta_t|m\rangle \ket n\bra m \label{A2}
\end{eqnarray}
and the scalar products above are given by the well known expressions of the coherent states in the Fock basis:
\be
\langle n|\gamma+\eta_t\rangle=e^{- |\gamma+\eta_t|^2/2}\frac{(\gamma+\eta_t)^n}{\sqrt{n!}}.
\ee
The strategy is now to use this relation and rewrite each sum in (\ref{A2}) as a coherent state but with a rescaled argument,
and then try to rewrite (\ref{A2}) as a single thermal state (with a different mean excitation number than $\varrho_0$).
To this end we note that:
\begin{eqnarray}
&&e^{-\frac{1}{2}|\gamma+\eta_t|^2}\sum_n\left(\frac{\bar n}{\bar n +1}\right)^{\frac{n}{2}}\frac{(\gamma+\eta_t)^n}{\sqrt{n!}}\ket n=\\&&e^{-\frac{1}{2}\frac{|\gamma+\eta_t|^2}{\bar n+1}}\left|\sqrt{\frac{\bar n}{\bar n +1}}(\gamma+\eta_t)\right\rangle.
\end{eqnarray}
Substituting this into (\ref{A2}) and reordering gives:
\begin{eqnarray}
&&\hat A_t 
=\frac{1}{\bar n +1}e^{-\frac{|\eta_t|^2}{1+2\bar n}}\int\frac{d^2\gamma}{\pi\bar n}e^{-\frac{1+2\bar n}{\bar n(\bar n +1)}
\left|\gamma+\frac{\bar n}{1+2\bar n}\eta_t\right|^2}\times\nonumber\\
&&\times \left|\sqrt{\frac{\bar n}{\bar n +1}}(\gamma+\eta_t)\right\rangle\left\langle\sqrt{\frac{\bar n}{\bar n +1}}(\gamma+\eta_t)\right|.\label{A3}
\end{eqnarray}
Note that since we are interested in $\tr\sqrt{\hat A_t}$ rather than $\hat A_t$ itself, there is a freedom 
of rotating $\hat A_t$ by a unitary operator, in particular by a displacement.
We now find such a displacement as to turn (\ref{A3}) into the thermal form.
Comparing the exponential under the integral in (\ref{A3}) with the thermal form \eqref{P-rep}, we see that the argument of the subsequent coherent states
should be proportional to $\gamma+ (\bar n \eta_t )/\left(1+2\bar n\right) $. Simple algebra gives:
\begin{eqnarray}
&&\left|\sqrt{\frac{\bar n}{\bar n +1}}(\gamma+\eta_t)\right\rangle\simeq\\
&&\hat D\left(\sqrt{\frac{\bar n}{\bar n +1}}\frac{\bar n+1}{1+2\bar n}\eta_t\right)\left|\sqrt{\frac{\bar n}{\bar n +1}}\left(\gamma+\frac{\bar n}{1+2\bar n}\eta_t\right)\right\rangle\nonumber,
\end{eqnarray}
where we have omitted the irrelevant phase factor  as we are interested in the projector on the above state.
Inserting the above relation into (\ref{A3}), dropping the displacements, and changing the integration variable:
$\gamma \to \sqrt{\bar n/\left(\bar n +1\right)}\left(\gamma+\left(1+2\bar n\right)\eta_t\right)$
gives:
\be
B[\Delta, t]=e^{-\frac{1}{2}\frac{|\eta_t|^2}{1+2\bar n}}\frac{1}{\sqrt{1+2\bar n}}\tr\sqrt{\varrho_{th}\left(\bar n^2/(1+2\bar n)\right)},
\ee
where $\varrho_{th}(\bar n)$ is the standard thermal state with the mean excitation number $\bar n$. We 
 use the Fock expansion \eqref{Fock} for $\varrho_{th}\left(\bar n^2/(1+2\bar n)\right)$ and obtain:
\begin{eqnarray}
&&B[\Delta, t]=e^{-\frac{|\eta_t|^2}{2+4\bar n}}\frac{1}{\sqrt{1+2\bar n}}\times\\
&&\times\left(1+\frac{\bar n^2}{1+2\bar n}\right)^{-\frac{1}{2}}
\sum_n \left(\frac{\bar n^2/(1+2\bar n)}{1+\bar n^2/(1+2\bar n)}\right)^{\frac{n}{2}}\\
&&=\exp\left[-\frac{1}{2}|\alpha(t)-\alpha'(t)|^2\text{th}\left(\frac{\hbar\beta\omega}{2}\right)\right],
\end{eqnarray}
where in the last step, we used the definition of $\eta_t$ from \eqref{UU} and $\bar n=1/(e^{\hbar\beta\omega}-1)$. 
Finally, using \eqref{alpha}, we obtain the desired result \eqref{B1}. It is interesting that the overlap factor looks very similar to the real part of the influence functional but with the inverse temperature dependence.

\section{Matsubara representation of the QFI kernel}\label{CLkernel}
Just like in the case of the noise kernel \cite{Breuer2002}, one can derive a formal analytic expression for the QFI kernel using the fermionic Matsubara representation \cite{Mahan2000}:
\ben
\mathrm{th} \left(\frac{\beta\hbar\omega}{2}\right) = \frac{4}{\beta \hbar \omega} \sum_{n=0}^{\infty}  \frac{1}{1+(\nu_n/\omega)^2},
\een   
with fermionic frequencies 
\be\label{nu_n}
\nu_n = \frac{(2n+1)\pi}{\hbar\beta}.
\ee
Substituting this into the QFI definition \eqref{qfik} and integrating term by term, we find:
\ben\label{phiseries}
&&\phi(\tau)=\frac{4M   \gamma \Lambda}{\hbar \beta} \sum_{n= 0}^{\infty} \frac{e^{-\Lambda |\tau|}-(\nu_n/\Lambda)\mathrm e^{-\nu_n |\tau|}}{1-(\nu_n/\Lambda)^2},
\een
which looks identical to the expansion of the noise kernel $\nu(\tau)$ \cite{Breuer2002}, except that instead of the bosonic frequencies $\nu_n = 2n\pi/(\hbar\beta)$ we now have the fermionic \eqref{nu_n}. In particular now $\nu_0=\pi/(\hbar\beta)\ne 0$ so that $\nu_n/\Lambda \gg 1$ for any $n$, including $n=0$. This complicates the analysis compared to the bosonic case, describing the influence functional, as there is now an interplay between $\Lambda$, $\nu_0$, and $\tau$.
The double integral in (\ref{Bmac}), which we denote by $\Phi[\Delta, t]$:
\begin{equation}\label{Phi1}
\Phi[\Delta, t] \equiv \int^t_0d\tau \int^\tau_0 d\tau'\Delta(\tau)\phi(\tau-\tau')\Delta(\tau')
\end{equation}
can be formally calculated term by term, using (\ref{phiseries}) and an explicit expression for the trajectories difference:
\begin{align}
&\Delta(\tau)\equiv X_{cl}(\tau)-X'_{cl}(\tau)\nonumber\\
&=\Delta X_0 \frac{\sin[\Omega(t-\tau)]}{\sin\Omega t}+\Delta X\frac{\sin\Omega\tau}{\sin\Omega t}.
\end{align}
We first slightly rearrange the expression (\ref{phiseries}):
\ben
&&\phi(\tau)=\\
&&\frac{4M   \gamma}{\pi} \sum_{n= 0}^{\infty} \frac{1}{(2n+1)}\frac{\Lambda^2}{1-(\Lambda/\nu_n)^2}\left(e^{-\nu_n|\tau|}-\frac{\Lambda}{\nu_n}e^{-\Lambda |\tau|}\right).\nonumber
\een
We now calculate term by term the double integral:
\begin{equation}
\Phi[\Delta, t] \equiv \int^t_0d\tau \int^\tau_0 d\tau'\Delta(\tau)\phi(\tau-\tau')\Delta(\tau'),
\end{equation}
using elementary integrals and obtain:
\ben
   && \Phi[\Delta, t]=\frac{4M\gamma}{\pi}\sum^\infty_{n=0} \frac{1}{(2n+1)}\frac{1}{1-(\Lambda/\nu_n)^2}\times \nonumber\\
    &&\frac{1}{\sin^2\Omega t} \Bigg[\Lambda^2\left[c_t(\nu_n)-(\Lambda/\nu_n)c_t(\Lambda)\right](\Delta X_0^2+ \Delta X^2)\nonumber\\
    &&+\Lambda^2\left[d_t(\nu_n)-(\Lambda/\nu_n)d_t(\Lambda)\right]\Delta X_0\Delta X\Bigg],\label{Phi2}
\een
with the coefficient defined as:
\ben
    &&c_t(\nu_n) =\label{c}\\
    &&\frac{1}{1+(\Omega/\nu_n)^2}\left[\frac{ t}{2\nu_n}-\frac{1}{4\nu_n\Omega}\sin(2\Omega t) -\frac{\sin^2\Omega t}{2\nu_n^2}\right]+\nonumber\\
   &&\frac{1}{[1+(\Omega/\nu_n)^2]^2}\left[\frac{\Omega^2}{\nu_n^4}-e^{-\nu_n t}\left(\frac{\Omega}{\nu_n^3}\sin\Omega t+\frac{\Omega^2}{\nu_n^4}\cos\Omega t\right)\right],\nonumber
\een

\ben
    &&d_t(\nu_n)=\label{d}\\
    &&\frac{1}{1+(\Omega/\nu_n)^2}\left[-\frac{t}{\nu_n}\cos\Omega t+\frac{1}{\Omega\nu_n}\sin\Omega t\right]-\nonumber\\
    &&\frac{1}{[1+(\Omega/\nu_n)^2]^2}\Bigg\{2\frac{\Omega^2}{\nu_n^4}\cos\Omega t+\nonumber\\
    &&e^{-\nu_n t}\left[\frac{\Omega^2}{\nu_n^4}+\left(\frac{\Omega}{\nu_n^2}\cos\Omega t+\frac{1}{\nu_n}\sin\Omega t\right)^2\right]\Bigg\}\nonumber
\een
and analogously for $c_t(\Lambda)$ and $d_t(\Lambda)$. The quantities that are small in the Caldeira-Leggett model are: 
\be\label{small}
\Lambda/\nu_n\ll 1, \ \Omega/\nu_n\ll 1, \ \Omega/\Lambda \ll 1,
\ee
which can be used to simplify the above expressions.

\bibliography{QBM2_v2}

\begin{thebibliography}{53}
\bibitem [{rem()}]{remark}%
  \BibitemOpen
  \href@noop {} {}\bibinfo {note} {In general, for the information encoding of
  the type $\varrho_i = U_i \varrho U_i^\dagger$, it holds that decoherence is
  a weaker process than the distinguishability, i.e. $|\Gamma_{ij}(t)|\leq
  B_{ij}(t)$. Let $\Gamma_{ij}(t)=\mathrm{tr}(\varrho U_i^\dagger
  U_j)\equiv\mathrm{tr} X$, where $X\equiv\sqrt\varrho U_i^\dagger U_j
  \sqrt\varrho$. Then $B_{ij}(t)=\mathrm{tr}(\sqrt{U_i\sqrt\varrho U_i^\dagger
  U_j \varrho U_j^\dagger U_i\sqrt\varrho
  U_i^\dagger})=\mathrm{tr}\sqrt{XX^\dagger}$. It is now an easy consequence of
  the polar decomposition that for any $X$, $|\mathrm{tr} X|\leq
  \mathrm{tr}\sqrt{XX^\dagger}$, from which $|\Gamma_{ij}(t)|\leq B_{ij}(t)$.
  However this algebraic relation does not say anything about the the type of
  decay of both quantities nor the dependence on the physical parameters and
  the existence of different lengthscales.}\BibitemShut {Stop}%
\end{thebibliography}

\end{document}